\documentclass{article}
\usepackage{amsmath}
\usepackage{amssymb}

\begin{document}

\begin{center}
\textbf{GRAVITATIONAL\ LENSING\ BY\ \ WORMHOLES}

\bigskip

Kamal K. Nandi$^{1,3,4,a}$, Yuan-zhong Zhang$^{2,3,b}$,

Alexander V. Zakharov$^{4,c}$

$^{1}$\textit{Department of Mathematics, University of North Bengal,
Darjeeling (W.B.) 743 430, India}

$^{2}$\textit{CCAST\ (World Laboratory), P.O.Box 8730, Beijing 100080,
Beijing, China}

$^{3}$\textit{Institute of Theoretical Physics, Chinese Academy of Sciences,
P.O. Box 2735, Beijing 100080, Beijing, China}

$^{4}$\textit{Joint Research Center for Mathematics and Physics (JRCMP),
Bashkir State Pedagogical University, 3-A, October Revolution Str., Ufa
450000, Russia}

\textit{\bigskip }

$^{a}$E-mail: kamalnandi1952@yahoo.co.in

$^{b}$E-mail: yzhang@itp.ac.cn

$^{c}$E-mail: zakharov@bspu.ru

\bigskip

\textbf{Abstract}
\end{center}

\textit{Gravitational lensing by traversable Lorentzian wormholes is a ew
possibility which is analyzed here in the strong field limit. Wormhole
solutions are considered in the Einstein minimally coupled theory and in the
brane world model. The observables in both the theories show significant
differences from those arising in the Schwarzschild black hole lensing. As a
corollary, it follows that wormholes with zero Keplerian mass exhibit
lensing properties which are qualitatively (though not quantitatively) the
same as those of a Schwarzschild black hole. Some special features of the
considered solutions are pointed out. }

\begin{center}
\bigskip
\end{center}

PACS number(s): 04.50.+h, 04.70.Bw, 95.30.Sf, 98.62.Sb

\begin{center}
\bigskip

\textbf{I. Introduction}
\end{center}

Gravitational lensing is an important and effective window to look for
signatures of peculiar astrophysical objects such as black holes (BH). This
field of activity has lately attracted a lot of interest among the physics
community. Early works focussed on the lensing phenomenon in the weak field
(for a review, see [1]), but weak field results can not distinguish between
various different solutions that are asymptotically flat. What one needs for
this purpose is a method of calculation in the strong field regime. Progress
in this direction have been initiated by Fritelli, Kling and Newman [2], and
by Virbhadra and Ellis [3]. However, the difficulty is that, in the strong
field, light deflection diverges at the photon sphere. (The conditions for
the existence of photon surfaces have been rigorously analyzed by Claudel,
Virbhadra and Ellis [4]). By an analytic approximation method, Bozza \textit{%
et al} [5] have shown that the nature of divergence of the deflection angle
becomes logarithmic as the light rays approach the photon sphere of a
Schwarzschild BH. This method has been successfully applied also in the
Reissner-Nordstr\"{o}m BH [6]. Virbhadra and Ellis [7] have further extended
the method of strong field lensing to cover the cases of Weak Naked
Singularity (WNS) and Strong Naked Singularity (SNS). Bozza [8] has
subsequently extended his analytic theory to show that the logarithmic
divergence near the photon sphere is a \textit{generic} feature for static,
spherically symmetric spacetimes. This work is a remarkable step forward in
the arena of gravitational lensing. Bhadra [9] has applied the procedure to
BHs in string theory. The extension of the strong field limit to Kerr BH has
also\ been worked out recently [10,11]. All these investigations have indeed
thrown up a richesse of information about the signatures of BH via lensing
mechanism.

There is however another exciting possibility that has not received enough
attention to date:\ It is lensing by stellar size traversable wormholes (WH)
which are just as interesting objects as BHs are. WHs have \textquotedblleft
handles\textquotedblright\ (throats) that connect two asymptotically flat
regions of spacetime and many interesting effects including light
propagation, especially in the Morris-Thorne-Yurtsever (MTY) WH spacetime,
have been extensively investigated in the literature [12]. WHs require
exotic matter (that is, matter violating at least some of the known energy
conditions) for their construction. The idea of this kind of matter has
received further justification in the notion of \textquotedblleft phantom
field\textquotedblright\ or \textquotedblleft dark matter" invoked to
interpret the observed galactic flat rotation curves or the present
acceleration of the Universe. Some works on lensing on a cosmological scale
involving dark matter do exist [13,14] but they have nothing to do with
Lorentzian WHs on a stellar scale. Nonetheless, it might be noted that
recent works by Onemli [15] show that the gravitational lensing by the dual
cusps of the caustic rings at cosmological distances may provide the
tantalizing opportunity to detect Cold Dark Matter (CDM) indirectly, and
discriminate between axions and weakly interacting massive particles
(WIMPs). It is also to be noted that local, static WH solutions threaded by
phantom matter have also been worked out recently [16].

Work in the direction of WH lensing has been initiated by Cramer \textit{et
al} [17] not very long ago and recently Safonova \textit{et al} [18] have
investigated the problem of lensing by negative mass WHs. A most recent work
by Tejeiro and Larra\~{n}aga [19] shows that Morris-Thorne type WHs
generally act like convergent lenses. Unfortunately, work on WH lensing, let
alone the strong field analysis, is still relatively scarce though
observables in WH lensing have the potential to serve a dual purpose: They
would establish not only the WH itself but also throw light on the existence
of classical exotic matter. This fact provides the basic motivation for the
present theoretical investigation.

We shall investigate the strong field lensing phenomenon in the WH solutions
belonging to the Einstein minimally coupled scalar \ field theory (EMS) as
well as the brane world model. (It is to be noted that novel effects of the
scalar field on gravitational lensing have been analyzed in Ref.[20] way
back in 1998 in the context of point like naked singularity lens.)
Importance of the EMS theory need not be repeated here. Suffice it to say
that it is the simplest scalar field theory. It can be connected to the
vacuum Brans-Dicke theory via the so-called Dicke transformations and to the
vacuum heterotic string theory. The static WH solutions in all these
theories have been well investigated [21-30]. On the other hand, brane
theory is a completely different proposition of great interest. The brane
world paradigm envisages that only gravity propagates in the 5-D bulk while
all other fields are confined to the 4-D brane. This idea leads to newer
models of local self-gravitating objects. It would therefore be interesting
to calculate the lensing effect in these models, especially in the strong
field limit.

Generically, the brane world BHs are far richer in structure than ordinary
BHs as they embody a synthesis of wormhole and black hole features. That is
why we refer to those objects here as WH/BHs. For instance, the effective
stress energy tensor could violate some of the energy conditions, though it
need not always be the case. This feature is not unexpected as the stress
tensor contains \textit{imprints} of the nonlocal free gravitational field
existing in the 5-D bulk which contributes negative energy [31]. Several
observable effects of the extra dimension on quasar luminosity in the
rotating models have been recently reported [32]. In the context of
spherical symmetry, the extra-dimensional bulk contribution essentially
implies a correction to the Schwarzschild solution but its horizon structure
remains undisturbed. The brane theory we have in mind is described by the
RS2 framework, that is, a single brane in a $Z_{2}$-symmetric 5-D
asymptotically anti-de Sitter bulk in which only gravity propagates while
all other fields are confined to the brane [33]. Strong field lensing in one
of the brane world BHs [34] have been carried out in Ref.[35] (and the weak
field lensing is calculated in Ref.[36]). Lensing in another class of brane
world BH (see below, Sec.IVA) has been investigated by Whisker [37]. Authors
in Refs.[35] and [37] have shown that such BHs could produce observables
that are significantly different from the Schwarzschild BH.

In this paper, we shall apply the strong field limit procedure, due to Bozza
[8], in the standard lensing (distinct from retrolensing) phenomenon by
static spherically symmetric WH solutions in the EMS theory and by the WH/BH
solutions in the brane theory. (Lensing in the weak field regime has been
investigated in Refs.[38].) We show that more spectacular differences can
appear in the observables in the strong field limit. This is our key result.

The paper is organized as follows. In Sec.IIA, we outline the procedure of
the strong field limit including the expressions for observables in Sec.IIB.
In Sec.IIIA, we deal with the lensing by a massive WH and in Sec.IIIB, with
the zero mass WH. In Sec.IVA, we point out that the brane world BH,
considered recently in Ref.[37] can also be interpreted as a self dual WH
harboring a globally strong naked singularity. Sec.IVB reveals
characteristic features of the strong field lensing by a different brane
world WH/BH solution. In Sec.V, we point out certain important aspects of
the considered WH solutions that should be useful in understanding the
lensing behavior. Finally, Sec.VI summarizes the results.

\begin{center}
\textbf{II. Strong Field Limit}

\textbf{A. Deflection angle}
\end{center}

We assume that the asymptotically flat spacetime describing a BH or WH is
centered at $L$ which serves as the lens. The observer $O$ and the source $%
S, $ which is to be lensed, are positioned in the flat region on either side
of $L$, but not necessarily along the same line. This is a plane
configuration of ordinary lensing, as distinct from retrolensing where both $%
O$ and $S$ are positioned only on one side of $L$. Let $I$ \ be the location
of the image of $S$ as observed by $O$ and that the extended $IS$ segment
meet the extended $OL$ segment at $X$. Defining the angles as $\langle
(OL,OS)=\beta $, $\langle (OL,OI)=\theta $, the lens equation follows from
the plane geometry [3]:
\begin{equation}
\tan \beta =\tan \theta -\frac{D_{LX}}{D_{OX}}\left[ \tan \theta +\tan
\left( \alpha -\theta \right) \right]
\end{equation}%
where $D_{OX}=D_{OL}+D_{LX}$ and $D_{PQ}$ is the Euclidean distance between $%
P$ and $Q$, and $\alpha $ is the deflection angle.

The generic spherically symmetric static metric for our purposes is (we take
$8\pi G=c=1$):
\begin{equation}
ds^{2}=A(x)dt^{2}-B(x)dx^{2}-C(x)d\Omega ^{2}
\end{equation}%
where $d\Omega ^{2}\equiv d\theta ^{2}+\sin ^{2}\theta d\varphi ^{2}$ is the
metric on a unit sphere. The photon sphere $x=x_{ps}$ is a solution of the
equation
\begin{equation}
\frac{C^{\prime }(x)}{C(x)}=\frac{A^{\prime }(x)}{A(x)}
\end{equation}%
in which the primes represent derivatives with respect to $x$. The impact
parameter $u$ is defined in terms of the closest approach distance $x=x_{0}$
as%
\begin{equation}
u=\sqrt{\frac{C(x_{0})}{A(x_{0})}}
\end{equation}%
The minimum impact parameter is given by%
\begin{equation}
u_{ps}=\sqrt{\frac{C(x_{ps})}{A(x_{ps})}}
\end{equation}%
From the equation of photon trajectory, it is easy to derive the deflection
angle
\begin{eqnarray}
\alpha (x_{0}) &=&-\pi +I(x_{0}) \\
I(x_{0}) &=&\int_{x_{0}}^{\infty }\frac{2\sqrt{B(x)}dx}{\sqrt{C(x)}\sqrt{%
\left( \frac{C(x)}{C(x_{0})}\right) \left( \frac{A(x_{0})}{A(x)}\right) -1}}
\end{eqnarray}

Bozza's procedure [8] for the strong field limit is based on the following
conditions: (a) The photon sphere $x=x_{ps}$ must exist, (b) The functions $%
A,B,C,A^{\prime },C^{\prime }$ must be positive for $x>x_{ps}$, (c) There
should exist a static limit, or horizon where $A(x_{s})=0.$ The last
condition is sufficient but not necessary. Then, define $y=A(x)$ , $%
y_{0}=A(x_{0})$ and
\begin{equation}
z=\frac{y-y_{0}}{1-y_{0}}
\end{equation}%
and rewrite the integral $I(x_{0})$ as%
\begin{equation}
I(x_{0})=\int_{0}^{1}R(z,x_{0})f(z,x_{0})dz
\end{equation}%
\begin{equation}
R(z,x_{0})=\frac{2\sqrt{By}}{CA^{\prime }}\left( 1-y_{0}\right) \sqrt{C_{0}}
\end{equation}%
\begin{equation}
f(z,x_{0})=\frac{1}{\sqrt{y_{0}-[(1-y_{0})z+y_{0}]\frac{C_{0}}{C}}}
\end{equation}%
where all functions without the subscript $0$ are evaluated at $%
x=A^{-1}[(1-y_{0})z+y_{0}]$. The function $R(z,x_{0})$ is regular for all
values of its arguments, but the function $f(z,x_{0})$ diverges as $%
z\rightarrow 0$ and it expands to second order like%
\begin{equation}
f(z,x_{0})\sim f_{0}(z,x_{0})=\frac{1}{\sqrt{\alpha _{1}z+\beta _{1}z^{2}}}
\end{equation}%
where the parameters $\alpha _{1},\beta _{1}$ depend on the closest approach
$x_{0}$ as%
\begin{equation}
\alpha _{1}=\frac{1-y_{0}}{C_{0}A_{0}^{\prime }}\left[ C_{0}^{\prime
}y_{0}-C_{0}A_{0}^{\prime }\right]
\end{equation}%
\begin{equation}
\beta _{1}=\frac{(1-y_{0})^{2}}{2C_{0}^{2}A_{0}^{\prime 3}}\left[
2C_{0}C_{0}^{\prime }A_{0}^{\prime 2}+\left( C_{0}C_{0}^{\prime \prime
}-2C_{0}^{\prime 2}\right) y_{0}A_{0}^{\prime }-C_{0}C_{0}^{\prime
}y_{0}A_{0}^{\prime \prime }\right]
\end{equation}%
Then the integral $I(x_{0})$ is resolved into a regular and a divergent part
and the latter gives the deflection angle to order $O(x_{0}-x_{ps})$ as%
\begin{equation}
\alpha (\theta )=-\overline{a}\ln \left( \frac{\theta D_{OL}}{u_{ps}}%
-1\right) +\overline{b}
\end{equation}%
where
\begin{equation}
\overline{a}=\frac{R(0,x_{ps})}{2\sqrt{\beta _{ps}}}
\end{equation}%
\begin{equation}
\overline{b}=-\pi +b_{R}+\overline{a}\ln \frac{2\beta _{ps}}{y_{ps}}
\end{equation}%
\begin{equation}
b_{R}=\int_{0}^{1}g(z,x_{ps})dz+O(x_{0}-x_{ps})
\end{equation}%
\begin{equation}
g(z,x_{ps})=R(z,x_{ps})f(z,x_{ps})-R(0,x_{ps})f_{0}(z,x_{ps})
\end{equation}%
\begin{equation}
\beta _{ps}=\beta _{1}\mid _{x_{0}=x_{ps}},y_{ps}=A(x_{ps})
\end{equation}
The function $g(z,x_{ps})$ is regular at $z=0$ [8].

\begin{center}
\textbf{B. Observables}
\end{center}

The relativistic images of the source are greatly demagnified in comparison
to weak field images because the photon trajectories wind several times
around the photon sphere before emerging outside. Yet, best results are
obtained when the source $S$, lens $L$ and the observer $O$ are highly
aligned. In this case we can assume that the angles $\theta $ and $\beta $
are small, but $\alpha =2n\pi +\Delta \alpha _{n}$, $n\in Z$ where $\Delta
\alpha _{n}$ is the residual angle after the trajectories wind the photon
sphere $n$ times. Under these conditions, the lens Eq.(1) reduces to

\begin{equation}
\theta =\beta +\frac{D_{LX}}{D_{OX}}\Delta \alpha _{n}
\end{equation}%
Defining $\alpha (\theta _{n}^{0})=2n\pi $, and using Eq.(15), we can write%
\begin{equation}
\theta _{n}^{0}=\frac{u_{ps}}{D_{OL}}(1+e_{n})
\end{equation}%
where%
\begin{equation}
e_{n}=e^{(\overline{b}-2n\pi )/\overline{a}}
\end{equation}%
The position $\theta _{n}^{0}$ and the magnification $\mu _{n}$ of the $n$th
relativistic image are:%
\begin{equation}
\theta _{n}=\theta _{n}^{0}+\frac{e_{n}u_{ps}(\beta -\theta _{n}^{0})D_{OX}}{%
\overline{a}D_{LX}D_{OL}}
\end{equation}%
\begin{equation}
\mu _{n}=\frac{1}{(\beta /\theta )\partial \beta /\partial \theta }\mid
_{\theta _{n}^{0}}\simeq \frac{e_{n}u_{ps}^{2}(1+e_{n})D_{OX}}{\overline{a}%
\beta D_{OL}^{2}D_{LX}}
\end{equation}

Now we bunch all the images together at $\theta _{\infty }=u_{ps}/D_{OL}$,
so that the outermost single image appears at $\theta _{1}$. Then define the
observables%
\begin{equation}
s=\theta _{1}-\theta _{\infty }
\end{equation}%
\begin{equation}
r=\frac{\mu _{1}}{\sum_{n=2}^{\infty }\mu _{n}}
\end{equation}%
which, respectively, are the separation and flux ratio between the bunch and
the outermost image. Using the relevant expressions, they simplify to%
\begin{equation}
s=\theta _{\infty }e^{(\overline{b}-2\pi )/\overline{a}}
\end{equation}%
\begin{equation}
r=e^{2\pi /\overline{a}}
\end{equation}%
We shall calculate the strong field coefficients $\overline{a}$, $\overline{b%
}$ and the observables $s$, $r$ for some physically interesting WH solutions
in the EMS and the brane world model.

\begin{center}
\textbf{III. EMS theory}
\end{center}

The field equations of the EMS theory are
\begin{equation}
R_{\mu \nu }=\kappa \Phi _{,\mu }\Phi _{,\nu }
\end{equation}%
\begin{equation}
\Phi _{;\mu }^{;\mu }=0
\end{equation}%
where $\Phi $ is the minimally coupled scalar field and $\kappa $ is a
constant free parameter. Note that the above equations are just the
conformally rescaled vacuum Brans-Dicke equations [24,30]. Clearly, all the
results in the sequel can be easily transcribed into those of Brans-Dicke
theory and further on, to string theory [29].

\begin{center}
\bigskip \textbf{A. Massive WHs}
\end{center}

\ \ A well known class of solutions of the EMS theory is the
Janis-Newman-Winnicour (JNW) [39] solution (or a variant of the Wyman\ [40]
solution):\
\begin{equation}
A(x)=\left( 1-\frac{2m}{x}\right) ^{\gamma },B(x)=\left( 1-\frac{2m}{x}%
\right) ^{-\gamma },C(x)=x^{2}\left( 1-\frac{2m}{x}\right) ^{1-\gamma }
\end{equation}%
\begin{equation}
\Phi (x)=\sqrt{\frac{1-\gamma ^{2}}{2\kappa }}\ln \left[ {1-}\frac{2m}{x}%
\right] \simeq \frac{q}{x}
\end{equation}%
\begin{equation}
\gamma =\frac{M}{m}
\end{equation}%
where $M$ is the ADM mass given by
\begin{equation}
M^{2}=m^{2}-\frac{1}{2}\kappa q^{2}
\end{equation}%
\begin{equation}
q=m\sqrt{\frac{2(1-\gamma ^{2})}{\kappa }}
\end{equation}%
is the asymptotic scalar JNW charge. In the field equations (30), we have
introduced a new constant parameter $\kappa $ that does not appear in the
observables but facilitates the analysis of the nature of the EMS solutions.
With a positive sign on the right hand side of Eq.(30), the stress tensor
represents ordinary scalar matter with positive energy density. The solution
(32,33) then has a globally strong naked singularity at $x=2m$ when $\gamma
<1$. However, with a negative sign on the right hand side, the stress tensor
represents energy condition violating exotic matter necessary for
constructing WHs. Now, this negative sign can be achieved in two ways:\ (i)
Take $\kappa =-2$ (that is, break all the energy conditions
\textquotedblleft by hand\textquotedblright\ or assume that this sign comes
as an input from another theory) and keep $\Phi $ real or (ii) Take $\kappa
=2$ but make $\Phi $ imaginary or which the same thing, $q$ imaginary. The
latter case also throws up a negative sign on the right side of Eq.(30) and
is completely physically valid as discussed by Armend\'{a}riz-P\'{\i}con
[41]. In either case, he solution represents the spacetime of a symmetric
traversable wormhole [22,24,27,28]. That there are two asymptotic regions
can be best seen by transforming the metric (30) into isotropic coordinates
via a radial transformation%
\begin{equation}
x=\rho \left( 1+\frac{m}{2\rho }\right) ^{2}
\end{equation}%
in which case the solution reduces to the Buchdahl solution [42] of 1959
given by
\begin{equation}
A(\rho )=\left( \frac{1-\frac{m}{2\rho }}{1+\frac{m}{2\rho }}\right)
^{2\gamma },B(\rho )=\left( 1-\frac{m}{2\rho }\right) ^{2(1-\gamma )}\left(
1+\frac{m}{2\rho }\right) ^{2(1+\gamma )},C(\rho )=\rho ^{2}B(\rho )
\end{equation}%
\begin{equation}
\Phi (x)=\sqrt{\frac{2(1-\gamma ^{2})}{\kappa }}\ln \left[ \frac{1-\frac{m}{%
2\rho }}{1+\frac{m}{2\rho }}\right] \simeq \frac{q}{\rho }
\end{equation}

The solution is invariant in form under radial coordinate transformation $%
\rho =\frac{m^{2}}{4\rho ^{\prime }}$ and hence one asymptotic region occurs
at $\rho =\infty $ and the other at $\rho ^{\prime }=0$, the two coordinate
patches meeting at $\rho =\rho ^{\prime }=\frac{m}{2}$. The WH throat occurs
at $\rho _{th}=\frac{m}{2}\left( \gamma +\sqrt{\gamma ^{2}-1}\right) $and
the requirement that $\rho _{th}$ be real and positive demands that $\gamma
>1$. This is the WH condition. The energy density $\rho _{D}$ and the scalar
curvature $R$ \ for the solutions (38) and (39) become
\begin{equation}
\rho _{D}=\left( \frac{1}{2}\right) \frac{m^{2}\left( 1-\gamma ^{2}\right) }{%
\left( 1-\frac{m^{2}}{4\rho ^{2}}\right) ^{2}}\left( \rho +\frac{m}{2}%
\right) ^{-2(1+\gamma )}\left( \rho -\frac{m}{2}\right) ^{-2(1-\gamma )}
\end{equation}%
\begin{equation}
R=2m^{2}\rho ^{4}\left( \rho +\frac{m}{2}\right) ^{-2(2+\gamma )}\left( \rho
-\frac{m}{2}\right) ^{-2(2-\gamma )}
\end{equation}%
Clearly, $\rho _{D}<0$ for $\gamma >1$ so that the Weak Energy Condition
(WEC) is violated. For $M\neq 0$, and for the case (i), we have $\gamma =%
\frac{M}{\sqrt{M^{2}-q^{2}}}$ and for the case (ii), defining $q=iq^{\prime }
$with $q^{\prime }>0$, we have $\gamma =\frac{M}{\sqrt{M^{2}-q^{\prime 2}}}$%
. Thus, $\gamma $ increases beyond unity if $q$ (or $q^{\prime }$) is
non-zero. The photon sphere appears at
\begin{equation}
\rho _{ps}=\frac{m}{2}\left[ 2\gamma \pm \sqrt{4\gamma ^{2}-1}\right]
\end{equation}%
It is clear that $\rho _{ps}>\rho _{th}$ so long as $\gamma >1$. All the
functions $A(\rho ),B(\rho ),C(\rho ),A^{\prime }(\rho )$ and $C^{\prime
}(\rho )$ are positive for $\rho >\rho _{ps}$. There is also the so-called
static limit at $\rho _{s}=\frac{m}{2}$ at which $A(\rho _{s})=0$. But the
surface $\rho _{s}=\frac{m}{2}$ is a strong naked singularity. However, $%
\rho _{ps}>\rho _{th}>\rho _{s}$ for $\gamma >1$ which implies that the
photon sphere hides the throat and the naked singularity. The situation
resembles the lensing scenario by Weakly Naked singularity (WNS) defined by
Virbhadra and Ellis [7] to the extent that the naked singularity is hidden
under the photon sphere. The occurrence of a throat hiding further the naked
singularity is a new feature in the present case. However, the main
difference is that the Virbhadra-Ellis choice of $\gamma $ is still less
than unity for $q>M$ since they defined $\gamma =\frac{M}{\sqrt{M^{2}+q^{2}}}
$.

The calculation of the strong field limit coefficients becomes awkward in
the isotropic coordinates and it is more convenient to use the metric (32)
which is in standard coordinates. Then the photon sphere appears at $%
x_{ps}=m(2\gamma +1)$. Without involving any loss of rigor, all that we need
to do is to take the WH range of $\gamma $ from from the foregoing analysis.
Now, in the case of Schwarzschild lensing, the value $u-u_{ps}=0.003$
involves an error of only $0.4\%$ from the exact position of the outer image
[8]. Taking this value as the starting point and using $u=\theta D_{OL}$,
the coefficients become
\begin{eqnarray}
\overline{a} &=&1 \\
\overline{b} &=&-\pi +b_{R}+\ln \frac{[(2\gamma +1)^{\gamma }-(2\gamma
-1)^{\gamma }]^{2}(2\gamma +1)}{2\gamma ^{2}(2\gamma -1)^{2\gamma -1}} \\
b_{R} &=&0.9496-0.1199(\gamma -1)+O(\gamma -1)^{2} \\
u_{ps} &=&\frac{(2\gamma +1)^{\gamma +\frac{1}{2}}}{2(2\gamma -1)^{\gamma -%
\frac{1}{2}}} \\
\beta _{ps} &=&\frac{[(2\gamma +1)^{\gamma }-(2\gamma -1)^{\gamma }]^{2}}{%
4\gamma ^{2}(4\gamma ^{2}-1)^{\gamma -1}} \\
\alpha (\gamma ) &=&-\overline{a}\ln \left( \frac{0.003}{u_{ps}}\right) +%
\overline{b}
\end{eqnarray}

It was shown in Ref.[8] that the deflection angle $\alpha (\gamma )$ \textit{%
decreases} from the Schwarzschild value in the range of naked singularity ($%
\gamma <1$). $\ $In contrast, the deflection angle $\alpha (\gamma )$
actually \textit{increases} from the Schwarzschild value with the value of
increasing $\gamma $ in the WH range ($\gamma >1$) as will be seen in the
Table I below. This behavior is markedly different from the case of naked
singularity or the Schwarzschild BH.

\begin{center}
\textbf{B. Massless WHs}
\end{center}

Consider a WH for which the ADM mass $M=0$. The cases (i) and (ii) mentioned
in Sec.IIIA respectively give $0=m^{2}+q^{2}$ and $0=m^{2}+q^{\prime 2}\ $%
which imply that both $m=0$ and $q=q^{\prime }=0$. This is a trivial case.
But we can \textit{also} have $M=m\gamma =0$ by putting $\gamma =0$, $m\neq
0 $. However, we must remember that, physically, the solar system tests fix $%
\gamma \sim 1$ while one is free to choose $m=0$ to achieve $M=0$. This
notwithstanding, we consider the reverse case ($\gamma =0$, $m\neq 0$) here
only as a mathematically interesting possibility. Thus, taking $\kappa =2$,
we have from Eq.(35) that%
\begin{equation}
q^{2}=m^{2}
\end{equation}%
implying that the gravitational stresses due to $m$ and non-gravitational
stresses due to $q$ exactly balance each other. This is an extremal
situation. Though $M=0$, and we should not expect any deflection at all, the
spacetime is \textit{not} flat. It is conceptually a classic example of
Wheeler's \textquotedblleft charge without charge\textquotedblright\ [43],
and it is a stable WH [41].

Due to our present choice of $\kappa $, $q^{2}<0$ as argued before, and so $%
m^{2}<0$. Let us take $m=-im^{\prime }$. Then, we have $q^{\prime
2}=m^{\prime 2}$ and moreover the energy density $\rho _{D}$ and scalar
curvature $R$ at the throat $\rho _{th}=\frac{m^{\prime }}{2}\ =\frac{%
q^{\prime }}{2}$ are given by
\begin{equation}
\rho _{D}=-\frac{1}{2q^{\prime 2}};R=-\frac{2}{q^{\prime 2}}
\end{equation}%
It would be interesting to analyze the effect of this massless curvature on
the light rays. Also, as the radial variable $\rho \rightarrow \infty $ in
Eqs.(40) and (41), both $\rho _{D}$ and $R\rightarrow 0$ implying that the
zero mass WH solution is asymptotically flat and also perfectly nonsingular
everywhere without a horizon. The solutions (38,39) become%
\begin{equation}
A(\rho )=1,B(\rho )=\left( 1+\frac{q^{\prime 2}}{4\rho ^{2}}\right)
^{2},C(\rho )=\rho ^{2}\left( 1+\frac{q^{\prime 2}}{4\rho ^{2}}\right)
^{2},\Phi (\rho )\simeq \frac{q^{\prime }}{\rho }
\end{equation}%
and the throat occurs$\ $at $\rho _{th}=\frac{q^{\prime }}{2}$. It also
represents the Ellis [44] \textquotedblleft drainhole\textquotedblright\
particle model. The Eqs.(51) can be expressed in proper distance $l=\rho -%
\frac{q^{\prime 2}}{4\rho }$ in a quite familiar form
\begin{equation}
ds^{2}=dt^{2}-dl^{2}-(l^{2}+q^{\prime 2})d\Omega ^{2},\Phi (l)=ArcTan\left(
\frac{l+\sqrt{l^{2}+q^{\prime 2}}}{q^{\prime }}\right)
\end{equation}%
where we have used the identity $ArcTan(x)=\frac{i}{2}\ln \left[ \frac{1-ix}{%
1+ix}\right] $ in $\Phi (\rho )$ of Eq.(39). The photon sphere exists and it
appears at $\rho _{ps}=\frac{q^{\prime }}{2}$ which coincides with the
nonsingular throat radius $\rho _{th}$. This is an extremal situation.
Looking at the necessary conditions, we see that (a) is satisfied even
though $A\neq 0$ anywhere. This only points to the fact that (c) is not a
necessary prerequisite. The condition (b) is marginally satisfied since all
the desired functions are positive and nonzero for $\rho >\frac{q^{\prime }}{%
2}$ except that $A^{\prime }=0$. [Note incidentally that in the standard
coordinates of the metric (1), the throat is at $x_{th}=m(1+\gamma )$ for $%
\gamma \neq 0$ and therefore $x_{th}=m$ for $\gamma =0$. The photon sphere
occurs also at $x_{ps}=m$, but the difficulty is that $C(x)\ngtr 0$ for $x>m$
in violation of the condition (b)]. Because $A^{\prime }=0$, the functions $%
R(z,x_{0})$ and $R(0,x_{m})$ diverge, and consequently do the coefficients $%
\overline{a}$ and $b_{R}$. \ Since the object is massless, there is no
possibility to consider the Schwarzschild lensing as a starting point in the
strong field analysis as we did before.

Perlick [45] has discussed detailed lensing properties of the Ellis
drainhole given by the metric (52). He discussed, in terms of an exact lens
equation, the cases that observer and light sources are (i) on different
sides and (ii) on the same side of the WH's throat. If the observer is
closer to the throat as the light source, the behavior of the bending angle
is similar in both the cases. In terms of the metric form (51), we can
integrate the deflection angle, Eq.(6), as%
\begin{equation}
\alpha (\rho )=-\pi +4ArcTanh\left( \frac{\rho }{\rho _{0}}\right)
\end{equation}
where $\rho =\rho _{0}$ is the closest approach. The function $\alpha (\rho )
$ is real only in the range $\left\vert \rho \right\vert <\rho _{0}$ and
this simply means that a light ray that starts from one asymptotic end and
passes through the throat of the WH can not go back to the same asymptotic
end from which it has started. It is very well possible that the light ray
goes to the other asymptotic end, as already discussed by Ellis [44]. [The
plot of $\alpha (\rho )$ $vs$ $\rho $ for $\rho _{0}=\rho _{th}=\frac{1}{2}$
(with units $q^{\prime }=1$) is same as the fig.8 in Ref.[45] derived
earlier.] Thus, there are two classes of light rays that start from one
asymptotic end:\ Members of the first class turn around before they reach
the neck, members of the second class pass through the throat and proceed to
the other asymptotic end. The borderline cases between the two are light
rays that asymptotically spiral towards the photon sphere at the throat.
These features are qualitatively similar to the light trajectories starting
at infinity in the Schwarzschild spacetime: Members of the first class turn
around before they reach the photon sphere, members of the second class pass
through the photon sphere and proceed to the horizon. The borderline
trajectories are those that spiral towards the photon sphere.

It is also possible to calculate the deflection angle, Eq.(6) using the
familiar proper form of the drainhole metric, Eq.(52), considered in
Ref.[45]. The minimum surface area $4\pi q^{\prime 2}$ appears at the throat
$l=0$ which is the same as $\rho _{ps}=\frac{q^{\prime }}{2}$. The extremal
situation $\rho _{th}=\rho _{ps}=\frac{q^{\prime }}{2}$ now translates into $%
l_{th}=l_{ps}=0$ Taking the closest approach at $l=l_{0}=a$ (say), we get,
for the exact deflection, an elliptic function
\begin{equation}
\alpha (a)=-\pi +\frac{2\sqrt{1+a^{2}}EllipticK\left[ -\frac{q^{\prime 2}}{%
a^{2}}\right] }{a}
\end{equation}
where $EllipticK(x)$ is a particular case of hypergeometric function. With
units in which $q^{\prime }=1$, we immediately find that $\alpha
(a)\rightarrow \infty $ (capture) as $a\rightarrow 0$. The plot $\alpha (a)$
$vs$ $a$ in the range $0<a<\infty $ again shows that the stable massless WH
acts qualitatively like a Schwarzschild deflector. What is interesting is
that the Keplerian mass $M$ is zero, yet light rays coming from the source
respond to this configuration and images the source. The strong deflection
limit around $a\sim 0$ in Bozza's formalism [8] is given by
\begin{equation}
\alpha (a)\simeq -\pi -2\ln (a-\frac{q^{\prime }}{2})+2\ln (2q^{\prime })
\end{equation}%
In terms of the distance $OL$, the same is given by $\alpha \simeq -\pi
-2\ln (OL)+2\ln (4q^{\prime })$. These plots approximate the exact
deflection pattern perfectly well.

\begin{center}
\ \ \ \ \ \ \ \ \ \ \ \textbf{IV. BH/WHs in the brane theory}$\ \ \ \ \ \ \
\ \ \ \ \ \ \ \ \ \ \ \ \ \ \ \ $
\end{center}

The 5-D Weyl tensor when projected onto the brane produces a trace-free
tensor $E_{\mu }^{\nu }$ appearing in the Shiromizu-Maeda-Sasaki brane field
equations [46]%
\begin{equation}
G_{\mu }^{\nu }=-\Lambda _{4}\delta _{\mu }^{\nu }-\kappa _{4}^{2}T_{\mu
}^{\nu }-\kappa _{5}^{4}\Pi _{\mu }^{\nu }-E_{\mu }^{\nu }
\end{equation}%
\begin{equation}
\Pi _{\mu }^{\nu }\equiv \frac{1}{2}[T_{\mu }^{\alpha }T_{\alpha }^{\nu
}-TT_{\mu }^{\nu }-\delta _{\mu }^{\nu }(T^{\alpha \beta }T_{\alpha \beta }-%
\frac{1}{2}T^{2})]
\end{equation}%
\begin{equation}
\Lambda _{4}\equiv \frac{1}{2}\kappa _{5}^{2}\left( \Lambda _{5}+\frac{1}{6}%
\kappa _{5}^{2}\lambda ^{2}\right)
\end{equation}%
\begin{equation}
\kappa _{4}^{2}\equiv 8\pi G_{N}\equiv \kappa _{5}^{4}\lambda /6\pi
\end{equation}%
where $G_{N}$ is the Newtonian Gravitational constant (we had earlier put $%
8\pi G_{N}=1$), $\Lambda _{4}$ and $\Lambda _{5}$ are, respectively, the 4-D
and 5-D cosmological constants, $\lambda $ is the brane tension. Visser and
Wiltshire [47] worked out an algorithm for finding solutions when matter
fields are present ($T_{\mu }^{\nu }\neq 0$) on the brane. To separate the
observable effects of pure bulk gravity from those due to ordinary matter on
the brane, we set $T_{\mu }^{\nu }=0$. As we are interested in the local
self-gravitating objects, we can ignore the cosmological $\Lambda _{4}$
term. The trace of the vacuum brane field equations then simply gives $R=0$
where $R$ is the Ricci scalar. This equation is solved to derive different
classes of brane world BH/WHs.

The general class of 4-D solutions is given by the metric of the form (2)
where $A(x)$ and $B(x)$ are two well behaved positive functions for $x>x_{h}$
and have a simple zero at $x=x_{h}$ defining the horizon. The singularities,
if any, of the BH solutions when propagated off the brane into the 5-D bulk
may make the AdS horizon singular (\textquotedblleft black cigar" [48]).
However, several classes of nonsingular, static, spherically symmetric
solutions of the brane world model have been proposed almost simultaneously
by Germani and Maartens [49] and by Casadio, Fabbri and Mazzacurati (GMCFM)
[50] and some quantum properties have also been investigated [51]. Under
certain assumptions on the behavior of the metric functions, Bronnikov,
Melnikov and Dehnen [52] have shown that the generic solutions can have $%
R\times S^{2}$ topology of spatial sections. Assuming asymptotic flatness at
large $x$, the global causal structure of such solutions coincides with a
section of the Kerr-Newman nonextremal solutions. We shall specifically
consider below two important GMCFM classes (I and II) of solutions.

\begin{center}
\textbf{A.\ GMCFM I solution}
\end{center}

In the units such that $2m=1$, the metric components are [49,50]:%
\begin{equation}
A(x)=\left( \varkappa +\lambda \sqrt{1-\frac{1}{x}}\right) ^{2},B(x)=\left(
1-\frac{1}{x}\right) ^{-1},C(x)=x^{2}
\end{equation}%
in which $\varkappa ,\lambda $ are arbitrary constants. Lensing in this
spacetime has already been investigated by Whisker [37], but some additional
observations seem to be in order. This is actually a self-dual solution of $%
R=0$ spacetimes with two asymptotic regions. For different domains of the
constants, this solution represents a Schwarzschild BH, naked singularity
and traversable wormholes [53]. Only for $\varkappa =0$ and $\lambda =1$, we
have a Schwarzschild BH. In isotropic coordinates $x=\rho (1+\frac{1}{4\rho }%
)^{2}$, the above metric becomes%
\begin{equation}
A(\rho )=\left[ \varkappa +\lambda \left( \frac{1-\frac{1}{4\rho }}{1+\frac{1%
}{4\rho }}\right) \right] ^{2},B(\rho )=\left( 1+\frac{1}{4\rho }\right)
^{4},C(\rho )=B(\rho )\rho ^{2}
\end{equation}%
The equation $A(\rho _{s})=0$ gives $\rho _{s}=\frac{1}{4}\frac{\lambda
-\varkappa }{\lambda +\varkappa }$. But at $\rho =\rho _{s}$, there appears
a naked singularity as can be seen from the following%
\begin{equation}
\rho _{D}=0,p_{\rho }=-\frac{2816\varkappa \rho ^{3}}{(1+4\rho )^{6}\sqrt{%
A(\rho )}},p_{\perp }=\frac{1408\varkappa \rho ^{3}}{(1+4\rho )^{6}\sqrt{%
A(\rho )}}
\end{equation}%
where $\rho _{D}$ is the density, $p_{\rho }$, $p_{\perp }$ are the radial
and cross radial components of pressure. The equation of state is that of
so-called \textquotedblleft dark radiation\textquotedblright\ given by $\rho
_{D}-(p_{\rho }+2p_{\perp })=0$. To get into Whisker's notation, one has
only to identify $\lambda =1+\epsilon $ and $\varkappa =-\epsilon $ so that $%
\rho _{s}=\frac{1+4\epsilon }{8}$. Due to the negative sign before $p_{\rho }
$, Averaged Null Energy Condition (ANEC), which is the weakest, is violated.
The metric (61) then represents a traversable symmetric WH with the throat
occurring at $\rho _{th}=\frac{1}{4}$. In order that $\rho _{th}>\rho _{s}$,
we must have $\epsilon <\frac{1}{4}$. This is the condition for
traversability though this condition is not strictly needed in the strong
field lensing calculation because the light rays are assumed to travel only
up to the photon sphere and not up to the throat. Since both the throat and
the naked singularity are hidden below $\rho _{ps}$, we have here a
situation like the EMS case investigated above.

However, it can be shown [27] that the total amount of ANEC violating matter
is $\Omega _{ANEC}=-\varkappa \ln \rho \mid _{\rho _{th}}^{\infty }$ which
diverges logarithmically with $\rho $. Therefore, unless some technical
modifications to the solution are made (e.g., as in Ref.[54]), the only way
to remove this divergence is to set $\varkappa =0$ which then produces the
trivial Schwarzschild BH solution. However, this divergence is not a problem
as the total gravitating mass is positive and finite as explicit
calculations will show in Sec.VC.

\begin{center}
\textbf{B. GMCFM II solution}
\end{center}

There another solution, described below, that also represents Schwarzschild
BH, naked singularity and traversable WHs which in the unit $2m=1$ has the
form [49,50].

\begin{equation}
A(x)=1-\frac{1}{x},B(x)=\frac{(1-\frac{3}{4x})}{(1-\frac{1}{x})(1-\frac{a}{x}%
)},C(x)=x^{2}
\end{equation}%
The nature of the solution depends on various choices of a constant
adjustable parameter $a$ interpreted as a bulk induced \textquotedblleft
tidal charge\textquotedblright\ - a Weyl tensor projection from the 5-D bulk
into the brane. The 4-D effective stress tensor components are:%
\begin{equation}
\rho _{D}=\frac{4a-3}{x^{2}(4x-3)^{2}},p_{x}=-\frac{4a-3}{x^{2}(4x-3)}%
;p_{\perp }=\frac{(2x-1)(4a-3)}{x^{2}(4x-3)^{2}}
\end{equation}%
For further interesting aspects of this spacetime, see Sec.V below. The
horizon appears at $x_{h}=1$. The spacetime structure depends on the
parameter $\eta =a-\frac{3}{4}$. Let us state the various cases [52]:\ (i)
If $\eta <0$ or $0<a<\frac{3}{4}$, the structure is that of a Schwarzschild
BH with a spacelike singularity at $x_{s}=\frac{3}{4}$. (ii) If $\eta >0$ or
$\frac{3}{4}<a<1$, then the solution describes a non-singular BH with a WH
throat at $x_{th}=a$. The causal structure is that of the ($1+1$)
dimensional subspace of a nonextremal Kerr BH solution. (iii) If $a=1$, then
we have a double horizon at $x_{h}=1$ with a timelike curvature singularity
at $x_{s}=\frac{3}{4}$. The global structure is that of an extreme
Reissner-Nordstr\"{o}m BH and finally (iv) The range $a>1$ corresponds to a
symmetric traversable WH with its throat occurring at either $x_{th}=1$ or $%
x_{th}=a$. For $a=\frac{3}{4}$, one recovers the Schwarzschild solution.

The photon sphere and the minimum impact parameter are given by%
\begin{equation}
x_{ps}=\frac{3}{2}
\end{equation}%
\begin{equation}
u_{ps}=\frac{3\sqrt{3}}{2}
\end{equation}%
which are independent of the tidal charge $a$. Thus, so long as $a\leq 1$ as
in the cases (i)-(iii), we see that the photon sphere covers the surfaces of
event horizon or singularities. In case (iv), too, the same situation occurs
if $1<a<\frac{3}{2}$. Thus, all physically meaningful solutions satisfying
the conditions (a)-(c) of Sec. IIA are contained in the range $0\leq a<\frac{%
3}{2}$. We can not take $a\geq \frac{3}{2}$ because in this case, the WH
throat radius $x_{th}=a$ exceeds that of the photon sphere $x_{ps}$. The
relevant coefficients work out to:%
\begin{equation}
\alpha _{1}=2-\frac{3}{x_{ps}}
\end{equation}%
\begin{equation}
\beta _{1}=\frac{3}{x_{ps}}-1
\end{equation}%
\begin{equation}
R(0,x_{ps})=\sqrt{\frac{3-4x_{ps}}{a-x_{ps}}}
\end{equation}%
\begin{equation}
\overline{a}=\left( \frac{1}{2}\right) \sqrt{\frac{x_{ps}(3-4x_{ps})}{%
(a-x_{ps})(3-x_{ps})}}
\end{equation}%
\begin{equation}
\overline{b}=-\pi +b_{R}+\frac{1}{2}\sqrt{\frac{x_{ps}(3-4x_{ps})}{%
(a-x_{ps})(3-x_{ps})}}\ln \left[ \frac{6-2x_{ps}}{x_{ps}-1}\right]
\end{equation}%
\begin{equation}
b_{R}=\int_{0}^{1}g(z,x_{ps})dz+O(x_{0}-x_{ps})
\end{equation}%
\begin{equation}
g(z,x_{ps})=-\frac{\sqrt{6}}{z\sqrt{3-2a}}+\frac{3\sqrt{2}}{z}\sqrt{\frac{z+1%
}{(3-2z)\left[ 3+2a(z-1)\right] }}
\end{equation}%
All the quantities above are well defined for $a<\frac{3}{2}$. The integral $%
b_{R}$ has no divergence on $[0,1]$ but its analytic evaluation in closed
form is rather unwieldy. However, we can easily expand $g(z,x_{ps})$ in
powers of $z$:%
\begin{equation}
g(z,x_{ps})=\frac{b_{1}(4a+3)}{(18-12a)}+\frac{b_{1}(64a-30a^{2}-33)}{%
8(3-2a)^{2}}z+O(z^{2})
\end{equation}%
where $b_{1}=3\sqrt{2}(9-6a)^{-\frac{1}{2}}$ which shows that $g(z,x_{ps})$
is perfectly regular at $z=0$. Since the solution under consideration
resembles that of Schwarzschild in many ways, especially, the photon sphere
appears exactly at the same value, we can, up to a good accuracy, consider
photon orbits for $u-u_{ps}=0.003$. \ We can then find the corresponding
value of $z$ by employing the expression [8]%
\begin{equation}
u-u_{ps}=c(x_{0}-x_{ps})^{2}
\end{equation}%
where $c$ is a constant. It turns out that%
\begin{equation}
c=\beta _{ps}\left[ \sqrt{\frac{A}{C^{3}}}\frac{C^{\prime 2}}{2(1-A)^{2}}%
\right] _{x=x_{ps}}=1
\end{equation}%
and so $x_{0}=1.554$. From the definition that $z=0$ at $x_{0}=x_{ps}$, we
can write%
\begin{equation}
z=\frac{A(x_{0})-A(x_{ps})}{1-A(x_{ps})}
\end{equation}%
which gives $z=z_{\min }=0.035$ corresponding to $u-u_{ps}=0.003$. By a
Taylor expansion around the Schwarzschild value $a=\frac{3}{4}$, we now
obtain%
\begin{equation}
b_{R}=\int_{z_{\min }}^{1}g(z,x_{ps})\mid _{a=\frac{3}{4}}dz+\left( a-\frac{3%
}{4}\right) \int_{z_{\min }}^{1}\frac{\partial g}{\partial a}\mid _{a=\frac{3%
}{4}}dz+O\left( a-\frac{3}{4}\right) ^{2}
\end{equation}%
Therefore
\begin{equation}
b_{R}\simeq 0.9496-\left( a-\frac{3}{4}\right) \times 1.565+O\left( a-\frac{3%
}{4}\right) ^{2}
\end{equation}%
The neglected higher order terms are smaller due to the gradually
diminishing factors in the powers of $\left( a-\frac{3}{4}\right) $ for $%
0\leq a<\frac{3}{2}$. The deflection $\alpha (x_{0})$ as a function of the
closest approach distance $x_{0}$ now works out to%
\begin{equation}
\alpha (x_{0})=b_{R}-\pi +\frac{1}{2}\Omega \ln \left[ \frac{(18-6x_{0})%
\sqrt{3(x_{0}-1)}}{2(x_{0}-1)\sqrt{x_{0}^{3}-3(x_{0}-1)}}\right]
\end{equation}%
where
\begin{equation}
\Omega \equiv \sqrt{\frac{x_{0}(3-4x_{0})}{(3-x_{0})(a-x_{0})}}
\end{equation}

Using the Schwarzschild value $\theta _{\infty }=16.87$ $\mu $
{\it arcsec}, the expressions for $\overline{a}$, $\overline{b}$,
$r$ , $s$ \ and $u$
as a function of closest approach $x_{0}$ turn out to be%
\begin{equation}
\overline{a}=\frac{\Omega }{2}
\end{equation}%
\begin{equation}
\overline{b}=b_{R}-\pi +\frac{\Omega }{2}\ln \left[ \frac{6-2x_{0}}{x_{0}-1}%
\right]
\end{equation}%
\begin{equation}
r=Exp\left[ \frac{4\pi }{\Omega }\right]
\end{equation}%
\begin{equation}
s=-\left[ \frac{33.74(x_{0}-3)}{x_{0}-1}\right] Exp\left[ \frac{2(b_{R}-3\pi
)}{\Omega }\right]
\end{equation}%
\begin{equation}
u=\sqrt{\frac{x_{0}^{3}}{x_{0}-1}}
\end{equation}%
From the above expressions, it is evident that, for the tidal charge value $%
a\simeq \frac{3}{2}$, the values for $\alpha (x_{0})$, $\overline{a}$,$%
\overline{b}$, $r$ and $s$ differ significantly from other choices of $a$
within the chosen range, especially near the photon sphere, $x_{0}\simeq
x_{ps}$. At $x_{ps}=\frac{3}{2}$, the relevant expressions become%
\begin{equation}
\overline{a}=\sqrt{\frac{3}{6-4a}}
\end{equation}%
\begin{equation}
\overline{b}=-\pi +2.123-1.565a+2.194\sqrt{\frac{1}{3-2a}}
\end{equation}%
\begin{equation}
r=Exp\left[ 2\pi \left( \sqrt{\frac{2}{3}(3-2a)}\right) \right]
\end{equation}%
\begin{equation}
s=101.22\times Exp\left[ \left( -7.301-1.565a\right) \sqrt{\frac{6-4a}{3}}%
\right]
\end{equation}%
Defining $u=\theta D_{OL}$, the deflection angle $\alpha (\theta )$ can be
rewritten as%
\begin{equation}
\alpha =-\overline{a}\ln \left( \frac{u-u_{ps}}{u_{ps}}\right) +\overline{b}%
+O(u-u_{ps})
\end{equation}%
that works out to%
\begin{equation}
\alpha (a)=-\pi +2.123-1.565a+10.478\sqrt{\frac{1}{3-2a}}
\end{equation}

The values of the observables are tabulated below. We see that the values of
$\overline{a}$,$\overline{b}$ continue to increase from the Schwarzschild
values ($\overline{a}=1,\overline{b}=-0.4009$) as we increase the tidal
charge. We also observe that the deflection angle $\alpha (a)$ \textit{%
increases} from the Schwarzschild value as the tidal charge is increased as
opposed to the \textit{decrease} caused by ordinary scalar fields (e.g., JNW
scalar field) [8]. This difference due to the tidal charge $a$ is
particularly manifest in the WH region corresponding to $1<a<\frac{3}{2}$.
For $a\sim \frac{3}{2}$, the deflection angle $\alpha (a)$ increases more
than three times compared to the value $\alpha (\frac{3}{4})$ for the
Schwarzschild BH. Such behavior could be interpreted as a signature for a WH
as well as effect of the extra dimension or tidal charge. The behavior of
the observables $r$ and $s$ too are very different from the Schwazschild BH
(or the JNW scalar field configuration) for different values of $a$,
especially at $a\sim \frac{3}{2}$.

\begin{center}
\textbf{Table I} \hspace{5mm}
\end{center}

\begin{center}
\begin{tabular}{c|c|c|c|c|c|c|c|c|c}
    \hline

 \multicolumn{1}{l|}
  {~} & \parbox[t]{1.1cm} {{\footnotesize $Sch$}: \\
   {\footnotesize $ \gamma =1$} \\
   {\footnotesize $a=\frac{3}{4}$} }
   & \parbox[t]{1.1cm} {~ \\  \footnotesize{ $\gamma =1.2$}}
   & \parbox[t]{1.1cm} {~ \\   \footnotesize{$\gamma =1.5$}}
   & \parbox[t]{1.1cm} {~ \\   \footnotesize{$ \gamma =1.8$}}
   & \parbox[t]{1.1cm} {~ \\    \footnotesize{$\gamma =2$}}
   & \parbox[t]{1.1cm} {~ \\ ~\\  \footnotesize{ $a=0.8$}}
   & \parbox[t]{1.1cm} {~\\ ~\\   \footnotesize{$a=1$}}
   & \parbox[t]{1.1cm} {~\\ ~\\   \footnotesize{$a=1.2$}}
   & \multicolumn{1}{l}{\parbox[t]{1.1cm} { ~\\ ~\\  \footnotesize{$a=1.4$}}}  \\
    \hline
 \parbox[t]{1cm}{$\alpha $ \footnotesize{ (\textit{rad.}) }}
 &6.36 &6.53 &6.72 &6.87 &6.96 &6.58 &7.89 &10.63 &20.22\\
  \hline
\parbox[t]{1.3cm}{$\theta _{\infty }$ \\
 \footnotesize{ ($\mu $ \textit{arcsec})} }
 &16.87 &20.56 &26.02 &31.36 &35.00 &16.87 &16.87 &16.87 &16.87\\
  \hline
\parbox[t]{1.3cm}{$s$ \\ \footnotesize{ ($\mu $ \textit{arcsec})}}
 &0.0211 &0.0205 &0.0197 &0.0189 &0.0185 &0.0261 &0.0726 &0.3047 &3.1618\\
  \hline
\parbox[t]{1cm}{$r_{m}$\\ \footnotesize{ (\textit{magn.})}}
 &6.82 &6.82 &6.82 &6.82 &6.82 &6.59 &5.57 &4.31 &2.50\\
  \hline
$u_{ps}$ &2.59 &3.16 &4.00 &4.82 &5.38 &2.59 &2.59 &2.59 &2.59\\
  \hline
$\overline{a}$ &1 &1 &1 &1 &1 &1.03 &1.22 &1.58 &2.74\\
  \hline
$\overline{b}$ &-0.4009 &-0.4292 &-0.4692 &-0.5073 &-0.5321
 &-0.4163 &-0.3895 &-0.0641 &1.6963\\
  \hline
\end{tabular}
\end{center}

\begin{center}
\textbf{V. Some features of the WH solutions}

\textbf{A. Massive EMS WH}
\end{center}

The EMS solutions (38,39) correspond to an equation of state $\rho
_{D}+p_{\rho }+2p_{\perp }=0$ for the WH case $\gamma >1$ since $\rho
_{D}=p_{\rho }$ and $-p_{\rho }=p_{\theta }=p_{\varphi }$. The equation is
the limiting case of the dark equation of state $p=w\rho $ where $w<-\frac{1%
}{3}$. (The \textit{phantom} equation of state is more stringent as it
requires $w<-1$ which is certainly not the case here.) The first observation
is that the total asymptotic gravitating mass $M=m\gamma $ is positive. It
can be calculated in various ways: by the ADM calculation [24] or from the
Einstein energy complex or even directly from the Eddington-Robertson
expansion of the centrally symmetric metric in isotropic coordinates [55]%
\begin{eqnarray}
ds^{2} &=&\left( 1-\frac{2\alpha _{1}M}{\rho }+\frac{2\beta _{1}M^{2}}{\rho
^{2}}+O(M^{3}/\rho ^{3})\right) dt^{2}-  \notag \\
&&\left( 1+\frac{2\gamma _{1}M}{\rho }+\frac{3\delta _{1}M^{2}}{2\rho ^{2}}%
+O(M^{3}/\rho ^{3})\right) [d\rho ^{2}+\rho ^{2}(d\theta ^{2}+\sin
^{2}\theta d\varphi ^{2})]
\end{eqnarray}%
\ Thus, the scalar field effect is already contained in the metric functions
$A(\rho )$, $B(\rho )$ in terms of $M=m\gamma $. This mass $M$ is the
gravitating mass and the test particles respond to it \textit{per se}; there
is in fact no way of measuring the bare $m$ if a scalar field
gravitationally couples to it. The Eddington-Robertson parameters for the
Buchdahl solution (38) are $\alpha _{1}=\beta _{1}=\gamma _{1}=1$, and the
post-PPN parameter $\delta _{1}=\frac{4}{3}-\frac{1}{3\gamma ^{2}}$. The
Buchdahl PPN parameters $\alpha _{1}$, $\beta _{1}$, $\gamma _{1}$ are
exactly the same as those in the Schwarzschild solution and at this level
EMS theory is indistinguishable from it. However, the deviation appears at
the post-PPN level and only finer and second order deflection measurements
can reveal the value of $\delta _{1}$. It is known that $\delta _{1}=1$ (or,
$\gamma =1$) corresponds to Schwarzschild solution while $\delta _{1}\neq 1$
would indicate a genuine deviation from it. The second order effect in
deflection (\textit{albeit }still in the weak field) can be easily
calculated by using the metric (93) involving $M$ and $\delta _{1}$ and
Eq.(6).

However, due to the nonlinearity of the field equations, the total amount of
WEC violating scalar matter $\Omega _{AWEC}$ in spacetime is slightly
different from $-q$ as a result of the generalized Gauss theorem in curved
spacetime. The exact difference can be seen from the volume AWEC integral,
which, for two sides of the WH becomes%
\begin{eqnarray}
\Omega _{AWEC} &=&2\times \left( \frac{1}{8\pi }\right) \int_{0}^{2\pi
}\int_{0}^{\pi }\int_{\rho th}^{\infty }\rho _{D}\sqrt{-g}\sin \theta d\rho
d\theta d\varphi   \notag \\
&=&-m(\gamma ^{2}-1)\ln \left( \sqrt{\frac{\gamma +1}{\gamma -1}}\right)
\simeq -m\sqrt{\gamma ^{2}-1}\left( 1-\frac{1}{2\gamma ^{2}}\right)
\end{eqnarray}%
As $\gamma \rightarrow 1$ (Schwarzschild case), $\Omega _{AWEC}\rightarrow 0$%
, as expected. Using the WH range, $\gamma >1$, we see that the $\Omega
_{AWEC}<0$. To express $\Omega _{AWEC}$ in terms of scalar charge $q$,
recall the two cases (i) and (ii) discussed in Sec. IIIA:\ If we take $%
\kappa =-2$, $\Phi $ real or $q=m\sqrt{\gamma ^{2}-1}$, then $\Omega
_{AWEC}\simeq -q\left( 1-\frac{1}{2\gamma ^{2}}\right) $, $q>0$ while for $%
\kappa =+2$, $\Phi $ imaginary or $q=im\sqrt{\gamma ^{2}-1}$, one has $%
\Omega _{AWEC}\simeq -iq\left( 1-\frac{1}{2\gamma ^{2}}\right) $. If we
integrate from $\rho _{ps}$ to $\infty $, we get $\Omega _{AWEC}\simeq
-q\left( 1-\frac{1}{8\gamma ^{2}}\right) $ and similarly the imaginary
version. The positive correction term proportional to $\frac{q}{\gamma ^{2}}$
slightly diminishes the quantity $\Omega _{AWEC}$ away from the value $-q$,
but this is due to nonlinear effects. For either of the two values of $%
\kappa $ and $q$, we see that $M^{2}=m^{2}+q^{2}=m^{2}\gamma ^{2}$. The
situation is the following: the WH is attractive and test particles, after
being pulled into the throat from both the mouths, attain zero acceleration
there. They can re-emerge into the other mouth by maintaining extra outward
radial accelerations from being pulled in again [56]. However, light rays
are captured at $\rho =\rho _{ps}$ but rays that pass close to it suffer
higher deflection angles than those due to Schwarzschild BH.

\begin{center}
\textbf{B. Massless EMS\ WH}
\end{center}

The zero mass WH with its metric given by Eqs.(51) or (52) is a stable
configuration (see Ref.[41] for details). The scalar field satisfies, to
first order in $\Phi (\simeq \frac{q^{\prime }}{\rho })$, a sourceless
equation $\frac{\partial }{\partial \rho }(\rho ^{2}\frac{\partial \Phi }{%
\partial \rho })=0$, yet the observers at a finite asymptotic $\rho $
measures a flux $4\pi q^{\prime }$ of the scalar charge though, in reality,
there is \textit{no} source. That is why the configuration is called a
\textquotedblleft charge without charge".

\begin{center}
\textbf{C. GMCFM\ I solution}
\end{center}

The GMCFM\ I solution is well discussed in the literature. Its asymptotic
physical mass can be found from the Einstein complex as follows [56]%
\begin{equation}
M=\underset{\rho \rightarrow \infty }{Lim}\frac{\rho A(B-1)}{\sqrt{2AB}}
\end{equation}%
which works out to $M=m(\varkappa +\lambda )$ which is finite and positive.
The WH is sustained entirely by the negative pressures as $\rho _{D}=0$.
With the identifications $\lambda =1+\epsilon $ and $\varkappa =-\epsilon $,
we immediately find that $M=m$. Therefore, \textit{energetically}, it is
still like the same Schwarzschild spacetime while, \textit{kinematically},
the null geodesics reveal that the strong field behavior is different from
that of the Schwarzschild, as the observables obtained by Whisker [37] show.

\begin{center}
\textbf{D. GMCFM II solution}
\end{center}

The GMCFM II solution exhibits certain remarkable features. Let us suspend
the unit $2m=1$ and restore $m$ for better comparison. First, we see that
the constituent matter is that of dark radiation given by $\rho
_{D}-(p_{x}+2p_{\perp })=0$. Second, it is impossible to ascertain the bulk
effect directly from $A(x)$ as it does not contain $a$. Therefore, we
proceed as follows. The integration of the Einstein complex of energy gives
the asymptotic physical mass of the solution [57]:%
\begin{equation}
M=\frac{1}{4}(2a+m)
\end{equation}%
which contains the bulk effect. From the Newtonian limit of $%
g_{00}(x)=A(x)=1-\frac{2m}{x}$, the Keplerian mass is always $m$ but the
asymptotic mass $M$, except in the special case $a=\frac{3}{2}m$, is
different. This feature is unlike the Buchdahl solution where both masses
are the same $M$ ($=m\gamma $). When we interpret Eq.(96) as a relation of
the type of Eq.(35)%
\begin{equation}
M^{2}=\frac{1}{16}\left( m^{2}+4a^{2}+4am\right)
\end{equation}%
we see that, unlike the EMS case, there is an extra interaction term $4am$
between the mass and the Weyl charge contributing to the total mass $M$. \
We see that the Weyl charge is additive to $m$, unlike the scalar charge.
Therefore, $a$ has the dimension of mass. Expressing $a$ in units of $M$
such that $a=\varepsilon M$, and using again the PPN expansion in standard
coordinates for a central mass $M$:

\begin{eqnarray}
ds^{2} &=&\left( 1-\frac{2\alpha _{1}M}{x}+\frac{2(\beta _{1}-\alpha
_{1}\gamma _{1})M^{2}}{x^{2}}+...\right) dt^{2}-  \notag \\
&&\left( 1+\frac{2\gamma _{1}M}{x}+\frac{4\delta _{1}M^{2}}{x^{2}}%
+...\right) dx^{2}-x^{2}(d\theta ^{2}+\sin ^{2}\theta d\varphi ^{2})
\end{eqnarray}%
and identifying the metric (63) with it, we find that $\alpha _{1}=\beta
_{1}=2(2-\varepsilon )$, $\gamma _{1}=1$, $\delta _{1}=\frac{1}{2}%
(2\varepsilon ^{2}-7\varepsilon +8)$. At $\varepsilon =\frac{3}{2}$, one
recovers the Schwarzschild values $\alpha _{1}=\beta _{1}=\gamma _{1}=$ $%
\delta _{1}=1$, as expected. When $a=\frac{3}{2}M$, we have $M=m$ from
Eq.(96). It turns out that first order tests can provide a value of $%
\varepsilon =\frac{3}{2}$ and can still be indistinguishable from GR, but
(weak field) second order deflection would check if $\delta _{1}=1$. If the
measured value of $\delta _{1}$ differs from unity, then the brane model of
stars would stand as a possible contender to that in the EMS theory.

The total amount of energy in spacetime due of bulk stress is
\begin{equation}
\Omega _{Total}=2\times \left( \frac{1}{8\pi }\right) \int_{0}^{2\pi
}\int_{0}^{\pi }\int_{0}^{\infty }\rho _{D}\sqrt{-g}\sin \theta dxd\theta
d\varphi =m-\sqrt{\frac{2am}{3}}
\end{equation}%
which is negative for the WH range $\frac{3m}{2}<a<3m$. (Recall that $%
x_{th}=a$ lies below $x_{ps}=3m$). The integral is zero either for $a=\frac{3%
}{2}m$ or, for $a=\frac{3}{2}M$ since, in both cases, $m=M$. The AWEC by
definition is
\begin{equation}
\Omega _{AWEC}=2\times \left( \frac{1}{8\pi }\right) \int_{0}^{2\pi
}\int_{0}^{\pi }\int_{x_{th}=a}^{\infty }\rho _{D}\sqrt{-g}\sin \theta
dxd\theta d\varphi =m
\end{equation}%
which is independent of $a$! This shows that the entire negative energy $-%
\sqrt{\frac{2am}{3}}$ is concentrated below the throat $0<x<a$. However,
there are some restrictions on the values of $a$. The pointwise WEC and the
AWEC are satisfied here for $a>\frac{3m}{2}$. For $a<\frac{3m}{2}$, WEC is
violated as $\rho _{D}<0$ but $\Omega _{AWEC}$ $=-i\infty $ which is
unphysical. The energetics of the brane model requires further study but we
see that the situation is very unlike the massive EMS WHs in which both WEC
and AWEC are violated for $\gamma >1$, the value of $\Omega _{AWEC}$ being
proportional to the scalar charge $-q$, as shown before.

Third, for $a<\frac{3m}{2}$, we find that the gravitating mass $M$ is
decreased from the Schwarzschild value $m$, \ the latter occurring at $a=%
\frac{3m}{2}.$ For $a>\frac{3m}{2}$, we find that $M>m$, which suggests that
the presence of the positive Weyl charge $a$ strengthens the attractive
force beyond that due to the Schwarzschild BH. This explains why there is an
enhancement in the deflection angle. The two surfaces $x_{ps}=3m$ and $%
x_{th}=a$ coincide when $a=3m$. At this extremal situation, there occurs
photon capture as the divergence in the deflection angle $\alpha (a)$ at $%
a=3m$ (which is the same as $a=\frac{3}{2}$) show in Eq.(92). At nonextremal
situations that we have considered, the throat lies below $x_{ps}$, and the
light rays do not reach the throat. We are interested, as mentioned before,
in the range $\frac{3m}{2}<a<3m$ and not in the range $a\geq 3m$ as, in this
case, it is the throat that covers the photon sphere, not the other way
around.

Finally, fourth, the WEC ($\rho _{D}\geq 0$) is locally preserved $\ $for $a>%
\frac{3m}{2}$ and so is AWEC which we saw to be independent of $a$. What
about the ANEC violation? Let us consider the volume ANEC integral%
\begin{eqnarray}
\Omega _{ANEC} &=&2\times \left( \frac{1}{8\pi }\right) \int_{0}^{2\pi
}\int_{0}^{\pi }\int_{x_{th}=a}^{\infty }(\rho _{D}+p_{x})\sqrt{-g}\sin
\theta dxd\theta d\varphi  \\
&=&-\int_{a}^{\infty }\frac{2(x-2m)(2a-3m)}{x^{2}(2x-3m)^{2}}\sqrt{-g}dx
\end{eqnarray}%
which contains only the radial pressure $p_{x}$. (We have not included the
transverse components of pressure as inequalities associated with $p_{\perp }
$ refer only to ordinary matter [54]). That is, the WH is maintained by
negative radial pressure [58]. Now, pointwise NEC violation $\rho
_{D}+p_{x}<0$ occurs when $\frac{3m}{2}<a<3m$, $2m<x<\infty $.
Unfortunately, the ANEC integral diverges logarithmically on $[a,\infty )$.
Such a divergence seems to be a generic feature of $R=0$ traversable WHs
[27]. However, it turns out that $\Omega _{ANEC}<0$ for $a<x\leq N$ where $N$
is any arbitrarily large but finite number. But as $N$ increases to
infinity, so does $\Omega _{ANEC}$ though not as rapidly. Thus, to have a
reasonable WH with \textit{finite} amount of ANEC violating matter, one
perforce needs to join the WH to the exterior vacuum Schwarzschild spacetime
at a certain value of the coordinate radius $x$, as is actually done with
self-dual $R=0$ WHs [54]. This peculiarity distinguishes the ANEC violation
due to the bulk effect from the violation due to sign reversed scalar field.
In the latter case, the $\Omega _{ANEC}$ tapers off smoothly at the
asymptotic region.

\begin{center}
\textbf{VI. Summary}
\end{center}

Gravitational lensing by WHs in the strong field limit is a new possibility
that has not been explored so far though pioneering theoretical works on
lensing by black holes or naked singularities exist in the recent
literature. On the other hand, \ WH solutions are physically important, many
of their properties have been widely discussed and applied to interpret
several outstanding problems in astrophysics [59-62]. It is thus only
natural that their analytic lensing properties be investigated as well.
Several static, spherically symmetric WH solutions are known, both in the
EMS theory and in the brane theory. Some of the brane world solutions
represent a synthesis of BH and WH spacetimes thereby providing a more
advanced and richer premise for the strong field lensing analysis. We have
undertaken a moderately comprehensive investigation here. Certain intrinsic
features of the lensing objects in question are also analyzed.

The sign reversed kinetic term in Eq.(30) yields regular, symmetric WH
solutions for the range of values $\gamma >1$ (The two options for $\kappa $
have been spelled out in the text.) For this range, the presence of the
scalar charge increases the Schwarzschild mass $m$, that is, the ADM mass $%
M=m\gamma >m$. The WH throat surface lies inside the photon sphere which has
been stipulated to play the limit of the strong field. That is, we have been
considering situations in which light rays pass very close to the exterior
of the photon sphere but obviously do not reach the throat. It was shown
that massive WHs in the EMS theory ($M\neq 0$) produce significantly
different values of deflection angles, and other observables as tabulated in
Table I. In contrast, in the case of naked singularity, $\Phi $ real, $%
\kappa >0$, $\gamma <1$, there is a decrease from the Schwarzschild mass,
that is, $M<m$. This explains why, in this case, the deflection angles
always show lesser values than those in the Schwarzschild case [8]. The
strong field lensing results thus show that the EMS scalar field exerts
stronger gravitational pull to light than that by the Schwarzschild BH. It
should however be remarked that an increase in ADM\ mass does not \textit{%
generally} imply an increase in the deflection angle. This follows from the
fact that the bending features are unaffected by a conformal factor whereas
a conformal factor does change the ADM mass.

The massless WH ($M=m\gamma =0$) corresponds to $\gamma =0$ but $m\neq 0$.
This is just a mathematically admissible possibility. The spacetime is
asymptotically flat at the two mouths. Lensing by these objects is
interesting due to the fact that it can reveal the presence of the geometric
curvature caused by the scalar field alone. It turns out that such
configurations also possess a photon sphere and behave like ordinary
deflectors.

The GMCFM I solution is treated as a brane world BH in the literature [37],
but it is actually a traversable WH. One recalls [52] that such class of
solutions can be a BH only when it is trivially Schwarzschild. Otherwise, it
is either a naked singularity or a WH. However, the throat radius $x_{th}=2m$
is hidden under the photon sphere justifying the application of the strong
field analysis. The GMCFM II solution has been investigated in detail here.
Though the minimum impact parameter $u_{ps}$ is exactly the same as that in
Schwarzschild BH, the spacetime itself is intrinsically very different.

Table I affords a comparison of the values of strong field observables for
WHs with those of Schwarzschild BH. It shows that the separation ($s$)
between the first image and the rest increases from that due to
Schwarzschild BH with increasing $\gamma $ ($>1$) but the increase is more
spectacular in the brane world WHs, especially in the region $a\sim 1.4.$
This suggests that the outermost image would be better visible in this case.
The flux ratio ($r_{m}$) or relative magnification always remains the same
in the EMS\ WHs but is considerably more than those in the brane world WHs.
These features are peculiar enough to observationally distinguish the
lensing sources under consideration.

Lensing phenomena in the WH environment offers a good possibility that one
might detect the presence not only of a WH, which is by itself interesting,
but also of the presence of naturally occurring exotic matter much advocated
on galactic or cosmological scales. VLBI observations of a clean system that
is devoid of intervention by accretion phenomena, can help us pick up the
right model or at least set limits on the observables. Still, it is not
unlikely that observations will favor just the usual Schwarzschild BH more
than any other advanced generalized solution considered here. Again, we note
that several astrophysical phenomena (like $\gamma $-ray bursts) can also be
explained by invoking new inputs (like negative energy fields or exotic
matter) [62], Thus, assuming that the center of our galaxy hosts, instead of
a BH, a WH threaded by exotic matter, situated at a distance $D_{OL}=8.5$ $%
kpc$ from the center of the Sun, then the angular position of the
set of relativistic images in the limit $n\rightarrow \infty $
would be $\theta _{\infty }\sim 17$ $\mu $ {\it arcsec}. We have
used this value as a basis for calculating $s$ in Table I.
However, due to considerable demagnification of relativistic
images, one would need a resolution of the order of $0.01$ $\mu $
{\it arcsec} and if this refinement is technologically attained in
future, then the observational limits can either accommodate or
rule out WH candidates as possible lensing sources.

\begin{center}
\textbf{Acknowledgments}
\end{center}

We are indebted to Valerio Bozza for suggesting several improvements on an
earlier version of the manuscript. We are thankful to Guzel N. Kutdusova and
Guo Cheng for useful assistance. This work is financially supported in part
by the TWAS-UNESCO program of ICTP, Italy, and by the Chinese Academy of
Sciences and in part by the National Basic Research Program of China \ under
Grant No. 2003CB716300 as well as by the NNSFC under Grant No. 90403032.

\bigskip

\textbf{References}

[1] P. Schneider, J. Ehlers, and E.E. Falco, \textit{Gravitational Lenses}
(Springer-Verlag, Berlin, 1992).

[2] S. Frittelli, T.P. Kling, and E.T. Newman, Phys. Rev. D \textbf{61},
064021 (2000).

[3] K.S. Virbhadra and G.F.R. Ellis, Phys. Rev. D \textbf{62}, 084003 (2000).

[4] C.-M. Claudel, K.S. Virbhadra, and G.F.R. Ellis, J. Math. Phys. \textbf{%
42}, 818 (2001).

[5] V. Bozza, S. Capozziello, G. Iovane, and G. Scarpetta, Gen. Relat. Grav.
\textbf{33}, 1535 (2001).

[6] E.F. Eiroa, G.E. Romero, and D.F. Torres, Phys. Rev. D \textbf{66},
024010 (2002); For strong field retrolensing, see: E.F. Eiroa and D.F.
Torres, Phys. Rev. D \textbf{69}, 063004 (2004)

[7] K.S. Virbhadra and G.F.R. Ellis, Phys. Rev. D \textbf{65}, 103004 (2002).

[8] V. Bozza, Phys. Rev. D \textbf{66}, 103001 (2002).

[9] A. Bhadra, Phys. Rev. D\textbf{\ 67}, 103009 (2003).

[10] V. Bozza, F. De Luca, G. Scarpetta, and M. Sereno, Phys. Rev. D \textbf{%
72}, 083003 (2005).

[11] S.E. V\'{a}squez and E.P. Esteban, Nuovo Cim. \textbf{119B}, 489 (2004).

[12] M.S. Morris and K.S. Thorne, Am. J. Phys. \textbf{56}, 395 (1988); M.S.
Morris, K.S. Thorne, and U. Yurtsever, Phys. Rev. Lett. \textbf{61}, 1446
(1988).

[13] S.M. Carroll, Living Rev. Rel. \textbf{4}, 1 (2001); R.R. Caldwell,
Phys. Lett. B \textbf{545}, 23 (2002).

[14] C. Charmousis, V. Onemli, Z. Qiu, and P. Sikivie, Phys. Rev. D \textbf{%
67}, 103502 (2003); Z.-H. Zhu, Int. J. Mod. Phys. D \textbf{9}, 591 (2000);
C.-M. Claudel, Proc. Roy. Soc. Lond. A \textbf{456}, 1455 (2000).

[15] V. Onemli, astro-ph/0510414.

[16] F.S.N. Lobo, Phys. Rev. D \textbf{71}, 084011 (2005); S.
Sushkov, Phys. Rev. D \textbf{71}, 043520 (2005).

[17] J.G. Cramer, R.L. Forward, M.S. Morris, M. Visser, G. Benford, and G.A.
Landis, Phys. Rev. D \textbf{51}, 3117 (1995).

[18] M. Safonova, D.F. Torres, and G.E. Romero, Phys. Rev. D \textbf{65},
023001 (2002).

[19] S.J.M. Tejeiro and R.E.A. Larra\~{n}aga, [gr-qc/0505054].

[20] K.S. Virbhadra, D. Narasimha, and S.M. Chitre, Astron. Astrophys.
\textbf{337}, 1 (1998).\

[21] A.G. Agnese and M. La Camera, Phys. Rev. D \textbf{51}, 2011 (1995).

[22] K.K. Nandi, A. Islam, and James Evans, Phys. Rev. D \textbf{55}, 2497
(1997).

[23] L.A. Anchordoqui, S.P. Bergliaffa, and D.F. Torres, Phys. Rev. D \textbf{%
55}, 5226 (1997)

[24] K.K. Nandi, B. Bhattacharjee, S.M.K. Alam, and J. Evans, Phys. Rev. D
\textbf{57}, 823 (1998).

[25] A. Bhadra and K. K. Nandi, Mod. Phys. Lett. A \textbf{16}, 2079 (2001).

[26] A. Bhadra and K.K. Nandi, Phys. Rev. D \textbf{64}, 087501 (2001).

[27] K.K. Nandi, Yuan-Zhong Zhang, and K.B.Vijaya   Kumar, Phys.
Rev. D \textbf{70},127503 (2004).

[28] K.K. Nandi, Yuan-Zhong Zhang, and K.B.Vijaya   Kumar, Phys.
Rev. D \textbf{70}, 064018 (2004).

[29] K.K. Nandi, Yuan-Zhong Zhang, Phys. Rev. D \textbf{70},
044040 (2004).

[30] Arunava Bhadra, Ion Simaciu, K.K. Nandi, and Yuan-Zhong
Zhang, Phys. Rev. D \textbf{71}, 128501 (2005).

[31] N. Dadhich, R. Maartens, P. Papadopoulos, and V. Rezania, Phys. Lett. B
\textbf{487}, 1 (2000).

[32] R. Rocha and C.H. Coimbra-Ara\'{u}jo, JCAP \textbf{0512}, 009 (2005)
(astro-ph/0510318, astro-ph/0509363); C.H. Coimbra-Ara\'{u}jo, R. Rocha, and
I.L. Pedron, Int. J. Mod. Phys. D \textbf{14}, 1883 (2005)
(astro-ph/0505132).

[33] L. Randall and R. Sundrum, Phys. Rev. Lett. \textbf{83}, 3370 (1999);
\textit{ibid}, \textbf{83}, 4690 (1999).

[34] R. Guedens, D. Clancy, and A.R. Liddle, Phys. Rev. D \textbf{66},
043513 (2002).

[35] E.F. Eiroa, Phys. Rev. D \textbf{71}, 083010 (2005).

[36] A.S. Majumder and N. Mukherjee, [astro-ph/0403405].

[37] R. Whisker, Phys. Rev. D \textbf{71}, 064004 (2005).

[38] D.F. Torres, E.F. Eiroa, and G.E. Romero, Mod. Phys. Lett. A \textbf{16}%
, 1849 (2001); A \textbf{16}, 973 (2001).

[39] A. Janis, E. Newman, and J. Winnicour, Phys. Rev. Lett. \textbf{20},
878 (1968);

[40] M. Wyman, Phys. Rev. D \textbf{24}, 839 (1985); R. Gautreau, Nuovo
Cimento B \textbf{62}, 360 (1969). For the equivalence of JNW, Wyman and
Buchdahl solutions, see: A. Bhadra and K.K. Nandi, Int. J. Mod. Phys. A
\textbf{16}, 4543 (2001).

[41] C. Armend\'{a}riz-P\'{\i}con, Phys. Rev. D \textbf{65}, 104010 (2002).

[42] H.A. Buchdahl, Phys. Rev. \textbf{115}, 1325 (1959).

[43] J.A. Wheeler, Phys. Rev. \textbf{48}, 73 (1957).

[44] H.G. Ellis, J. Math. Phys. \textbf{14}, 104 (1973); \textbf{15},
520(E)\ (1974). For scattering problems in such spacetimes, see: G. Cl\'{e}%
ment, Int. J. Theor. Phys. \textbf{23}, 335 (1984); L. Chetouani and G. Cl\'{%
e}ment, Ge. Relat. Gravit. \textbf{16}, 111 (1984).

[45] V. Perlick, Phys. Rev. D \textbf{69}, 064017 (2004).

[46] T. Shiromizu, K.I. Maeda, and M. Sasaki, Phys. Rev. D
\textbf{62}, 024012 (2000).

[47] M. Visser and D.L. Wiltshire, Phys. Rev. D \textbf{67}, 104004 (2003).

[48] S.W. Hawking, G.T. Horowitz, and S.F. Ross, Phys. Rev. D \textbf{51},
4302 (1995).

[49] C. Germani and R. Maartens, Phys. Rev. D \textbf{64}, 124010 (2001).

[50] R. Casadio, A. Fabbri, and L. Mazzacurati, Phys. Rev. D
\textbf{65}, 084040 (2002).

[51] R. Casadio and B. Harms, Phys. Lett. B \textbf{487}, 209 (2000).

[52] K.A. Bronnikov, V.N. Melnikov, and H. Dehnen, Phys. Rev. D \textbf{68},
024025 (2002).

[53] N. Dadhich, S. Kar, S. Mukherjee, and M. Visser, Phys. Rev. D \textbf{65%
}, 064004 (2002).

[54] M. Visser, S. Kar, and N. Dadhich, Phys. Rev. Lett. \textbf{90}, 201102
(2003).

[55] S. Weinberg, \textit{Gravitation and Cosmology}, (John Wiley, New York,
1972), p. 183

[56] L.H. Ford and T.A. Roman, Phys. Rev. D \textbf{53}, 5496 (1996).

[57] K.S. Virbhadra, Phys. Rev. D \textbf{60}, 104041 (1999).

[58] K.A. Bronnikov and S.-W. Kim , Phys. Rev. D \textbf{67}, 064027 (2003).

[59] Ke-Jian Jin, Yuan-Zhong Zhang, and Zong-Hong Zhu, Phys. Lett.
\textbf{A 264}, 335 (2000).

[60] D. Hochberg and T.W. Kephart, Phys. Rev. Lett. \textbf{70}, 2665
(1993); Phys. Lett. B \textbf{268}, 377 (1991); Gen. Relat. Gravit. \textbf{%
26}, 219 (1994).

[61] D.H. Coule, Phys. Lett. B \textbf{450}, 48 (1999); D.H. Coule and K.
Maeda, Class. Quant. Gravit. \textbf{7}, 955 (1990).

[62] V. Frolov and I. Novikov, Phys. Rev. D \textbf{48}, 1607 (1993).

[63] F. Hoyle, G. Burbidge, and J.V. Narlikar, Proc. Roy. Soc. Lond. \textbf{%
A 448}, 191 (1995); Mon. Not. Roy. Astron. Soc. \textbf{267}, 1007 (1994).

\bigskip

\end{document}